\documentclass{sigchi}
\usepackage{siunitx}
\usepackage{float}
\usepackage[show]{chato-notes} 
\usepackage{amsmath}
\usepackage[titlenumbered,ruled]{algorithm2e}
\usepackage{algpseudocode}
\usepackage{pifont}
\usepackage{times}
\usepackage{tabulary}
\usepackage{booktabs}
\usepackage{multirow}
\usepackage{balance}  
\usepackage{graphics} 
\usepackage[T1]{fontenc}
\usepackage{txfonts}
\usepackage{mathptmx}
\usepackage[pdftex]{hyperref}
\usepackage{color}
\usepackage{booktabs}
\usepackage{textcomp}
\usepackage{subfigure}
\usepackage{microtype} 
\usepackage{ccicons}  
\usepackage[utf8]{inputenc} 

\def\plaintitle{}

\def\emptyauthor{}
\def\plainkeywords{}

\makeatletter
\def\url@leostyle{%
  \@ifundefined{selectfont}{
    \def\UrlFont{\sf}
  }{
    \def\UrlFont{\small\bf\ttfamily}
  }}
\makeatother
\def\pprw{8.5in}
\def\pprh{11in}

\definecolor{linkColor}{RGB}{6,125,233}
\hypersetup{%
  pdftitle={\plaintitle},
  pdfauthor={\emptyauthor},
  pdfkeywords={\plainkeywords},
  bookmarksnumbered,
  pdfstartview={FitH},
  colorlinks,
  citecolor=black,
  filecolor=black,
  linkcolor=black,
  urlcolor=linkColor,
  breaklinks=true,
}

\begin{document}

\title{Linguistic Diversities of Demographic Groups in Twitter%
\titlenote{\textcolor{red}{\textbf{This is a pre-print of a paper accepted to appear at HyperText 2017}}.\\ \\
\textcolor{blue}{Linguistic Diversities of Demographic Groups in Twitter.\\
Pantelis Vikatos, Johnnatan Messias, Manoel Miranda, and Fabricio Benevenuto\\
In Proceedings of the 28th ACM Conference on Hypertext and Social Media (HT'17). Prague, Czech Republic. July 2017.\\\\
\textbf{Bibtex: \url{http://johnnatan.me/bibtex/vikatos_ht_2017.bib}\\
For more published work: \url{http://johnnatan.me}}}}}

\numberofauthors{4}
\author{%
\alignauthor
Pantelis Vikatos\\
       \affaddr{University of Patras}\\
       \affaddr{Rio, Greece}\\
       \email{vikatos@ceid.upatras.gr}
\alignauthor
Johnnatan Messias\\
       \affaddr{Universidade Federal de Minas Gerais}\\
       \affaddr{Belo Horizonte, Brazil}\\
       \email{johnnatan@dcc.ufmg.br}
\alignauthor
Manoel Miranda\\
       \affaddr{Universidade Federal de Minas Gerais}\\
       \affaddr{Belo Horizonte, Brazil}\\
       \email{manoelrmj@dcc.ufmg.br}
\alignauthor
Fabricio Benevenuto\\
       \affaddr{Universidade Federal de Minas Gerais}\\
       \affaddr{Belo Horizonte, Brazil}\\
       \email{fabricio@dcc.ufmg.br}
}

\maketitle

\begin{abstract}

The massive popularity of online social media provides a unique opportunity for researchers to study the linguistic characteristics and patterns of user's interactions. In this paper, we provide an in-depth characterization of language usage across demographic groups in Twitter.
In particular, we extract the gender and race of Twitter users located in the U.S. using advanced image processing algorithms from Face++. Then, we investigate how demographic groups (i.e. male/female, Asian/Black/White) differ in terms of linguistic styles and also their interests. We extract linguistic features from $6$ categories (affective attributes, cognitive attributes, lexical density and awareness, temporal references, social and personal concerns, and interpersonal focus), in order to identify the similarities and differences in particular writing set of attributes. In addition, we extract the absolute ranking difference of top phrases between demographic groups. As a dimension of diversity, we also use the topics of interest that we retrieve from each user. Our analysis unveils clear differences in the writing styles (and the topics of interest) of different demographic groups, with variation seen across both gender and race lines. We hope our effort can stimulate the development of new studies related to demographic information in the online space.

\end{abstract}

\section{Introduction} \label{sec:introduction}

The number of users in online social networking sites, such as Facebook and Twitter increases each day. As of the third quarter of 2016, Facebook and Twitter have 1.79 billion\footnote{\url{http://www.statista.com/statistics/264810/number-of-monthly-active-facebook-users-worldwide/}} and 317 million\footnote{\url{http://www.statista.com/statistics/282087/number-of-monthly-active-twitter-users/}} monthly active users, respectively, sharing content about their daily lives and things that happen around them. This massive popularity of online social media provides the opportunity to detect useful characteristics and patterns about users and their interconnections. For instance, patterns are valuable for marketing and advertisement companies which capture users' behavior and needs in order to promote products, specifically on a target group. In terms of group, demographics constitute a significant factor to cluster people and understand their behavior. 
Twitter provides a plethora of different information, e.g. posts, social connections. However, it lacks data about demographics such as gender, race, or age. We deal with this absence of this information using profile image as an input of deep learning algorithms on image processing.
We are interested in extracting demographic status in a large scale and correlate it with available information on the social media.
Twitter is a micro-blogging platform so the main way of communication and action is by posting texts (tweets). The use of natural language processing in these type of data can extract many features describing cognitive and user' personal concerns. 


Many studies have used text analysis to study the user behavior in the online space~\cite{icwsm10cha,benevenuto@imc09,cha@tsmca12,correa-2015-anonymityShades}. Our work provides a complementary perspective to these efforts, by providing a characterization of language usage (i.e. common phrases and topics of interest), but grouping users according to their gender and race. Our effort is motivated by previous studies that uses computational linguistics in order to extract patterns about demographic information~\cite{DeChoudhury:2017:GCD:2998181.2998220}, but our effort further explores race as a new demographic dimension. Our findings reveal significant differences between the linguistic
content shared by female and male users as well as Asian, Black, and White and can be used for automatically categorization of Twitter users through their texts.

The main challenge is that users in Twitter are prone not to provide information about demographics. In our work, we crawled a large scale sample of active Twitter users and then we identify the gender and race of about $1.6$ million users located in U.S by using Face++\footnote{\url{http://www.faceplusplus.com}}~\cite{fan2014learning,yin2015learning}, a face recognition software able to recognize gender and race of identifiable faces in the user's profile pictures. Actually, the state of the art algorithms, for pattern recognition and image processing, can provide with high accuracy the gender, race, and even the age of an individual via his/her image. 
From the demographic recognized users, we gathered tweets of $304,477$ users to characterize linguistic patterns. Particularly, we extract the absolute ranking difference of top phrases between demographic groups. As a dimension of diversity, we also use the topics of interest that we retrieve from each user. Our analysis concludes that there are clear differences in the way of writing across different demographic groups in both gender and race domains as well as in the topic of interest.


The rest of the paper is organized as follow. Section \ref{sec:related} provides a review of the relevant literature. Then, Section \ref{sec:dataset} presents the Twitter and demographic dataset. After that, the analysis and discussion of linguistic differences and topic of interests are presented. Finally, the last section summarizes our results and offers some concluding remarks.

\section{Related Work} \label{sec:related}
In this section, we review the related literature along two axes. First, we discuss the methodology used by efforts that measure demographic factors in Twitter. Then, we refer to studies that combine linguistic with demographic status.

\subsection{Demographics in Social Media} 

One of the first efforts to extract and analyze demographic information presents a comparative study between the demographic distribution of gender/race of Twitter users and U.S. population \cite{mislove2011understanding}.
After that, several efforts have arisen that investigate demographic information, in various social media, using different strategies for distinct purposes \cite{blevins2015jane,burger2011discriminating,karimi2016,hannak2017@icwsm,hannak-2017-cscw}. Particularly, in terms of text analysis, Cunha \textit{et al.}~\cite{Cunha:2012:GBS:2309996.2310055} used Twitter data to analyze the difference between males and females in terms of generation of hashtags. Their results emphasize gender as  factor able to influence the user's choice of specific hashtags to a specific topic.

Recent studies focused on demographics~\cite{blevins2015jane,karimi2016,liu2013s,reis2017,nilizadeh2016twitter} present methodologies to extract the necessary data through analysis and pattern matching of screen/full name as well as descriptions of user profiles and image in the profile status. Particularly, Chen \textit{et al.}~\cite{chen2015comparative} focus on demographic inference using namely profile self-descriptions and profile images. They categorize demographic status using as signals users' names, self-descriptions, tweets, social networks, and profile images to infer attributes as ethnicity, gender, and age. 
An alternative approach, Culotta \textit{et al.}~\cite{culotta2015predicting} declare that the demographic profiles of visitors to a website are correlated with the demographic profiles of followers of that website on the social network and propose a regression model to predict demographic attributes such as gender, age, ethnicity, education, income, and child status.
More recently, An \textit{et al.}~\cite{an2016greysanatomy} provide an accurate scheme in order to predict gender and race using the correlation of hashtags that are used in different demographic groups.

Finally, our effort uses the similar strategy to gather demographic information as Chakraborty \textit{et al.}~\cite{chakraborty2017@icwsm}, but we investigate very different research questions as we focus on the linguistic analysis of demographic groups.


\subsection{Demographics and linguistic analysis} 


In the field of demographics, most studies use linguistic analysis in order to extract useful features for predicting demographic information as gender, race, and age.
Burger \textit{et al.}~\cite{burger2011discriminating} produce n-grams from users' tweets, description, screen name, and full name, in order to predict Twitter user gender. They conclude that the training of an SVM classifier with the combination of all factors can create an efficient and accurate prediction scheme ($92\%$ acc) for gender classification.
Also, Chen \textit{et al.}~\cite{chen2015comparative} introduce a similar methodology for predicting gender, ethnicity, and age. However, using n-grams from the social neighbors, including followers and friends, and the distribution of 100 generated topics of LDA algorithm as the input of SVM classifier. 
Their results present that the performance of classification is much lower in terms of ethnicity and age. 
Gilbert \textit{et al.}~\cite{Gilbert:2013:INT:2470654.2481336}
present an interesting statistical overview in Twitter and Pinterest using textual analysis and comparing what users text on Pinterest to what they text on Twitter.

We mainly motivate our research based on Choudhury \textit{et al.}~\cite{DeChoudhury:2017:GCD:2998181.2998220} study which discover gender and cultural differences in Twitter. They correlate several linguistic features to mental illness. Our findings reinforce their observations about linguistic and topical differences against male and female users in Twitter and also contribute with a new analysis of race.

\section{Demographic Information Dataset} \label{sec:dataset}

This section focuses on the procedure of data collection in order to extract useful inference about the discrimination of demographic status of a Twitter user. Our ultimate goal consists of gather demographic characteristics as gender and race as well as attributes about social behavior and tweet activity of active U.S. Twitter users. Next, we describe our steps to create this dataset and also discuss its main limitations.

\subsection{Twitter dataset gathered}
Our procedure uses the provided information from Twitter Stream API~\footnote{\url{http://dev.twitter.com/streaming/public}} in order to identify active Twitter users. We use a time window of three complete months from July to September 2016, collecting $341,457,982$ tweets posted by $50,270,310$ users.

Due to the fact that geographic coordinates are available on Twitter only for a limited number of users (i.e. $<$ $2\%$)~\cite{icwsm10cha}, our strategy to identify U.S. Twitter users is based on the time zone information to retrieve users which are actually from the US as the methodology in previous efforts~\cite{chakraborty2017@icwsm,kulshrestha2012geographic} presented.

We filtered users that provided free text location indicating they are not U.S. (i.e. Montreal, Vancouver, Canada). We end up with a dataset containing $6,286,477$ users likely located in the United States.  

\subsection{Crawling Demographic Information}

The field of demographic status is not mandatory when a user registers in Twitter and, thus, the direct retrieval of gender, race, or even age is not feasible. There are several studies related to demographic information in Twitter that attempt to infer the user's gender from the user name~\cite{blevins2015jane,karimi2016,liu2013s,mislove2011understanding}.
Also, some works use pattern based methodology to identify age~\cite{sloan2015tweets} in Twitter profile description using regular expressions {\it `$25$ yr old'} or {\it `born in $1990$'}.

Here, we use a different strategy that allows us to extract the demographic dimension using the profile picture of each user. To do that, we needed to gather the profile picture web link of all Twitter users identified as located within the United States. In December $2016$, we crawled the profile picture's URLs of about $6$ million users, discarding $4,317,834$ ($68.68\%$) of them.  We discarded users in two situations, first when the user does not have a profile picture and second when the user has changed her picture since our first crawl. When users change their picture, their profile picture URL changes as well, making it impossible for us to gather these users in a second crawl.   

From the remaining $1,968,643$ users, we submitted the profile picture web links into the {\it Face++ API}. Face++ is a face recognition platform based on deep learning~\cite{fan2014learning,yin2015learning} able to identify the gender (i.e. male and female) and race (limited to Asian, Black, and White) from recognized faces in images.

\begin{table}[h]
 \centering
\caption{Dataset construction}
\label{table:dataset}
\begin{tabular}{| c | c | p{3.3cm} |}
\hline
\textbf{Phase} & \textbf{Number of Users}\\
\hline
Crawling $3$ months of Tweets & $50$ million\\
\hline
Filtering U.S. users &  $6$ million\\
\hline
U.S. users with profile image &  $2$ million\\
\hline
U.S. users with one face (Baseline) &  $1.6$ million\\
\hline
U.S. users with crawled tweets  &  $304$ thousand \\
\hline
\end{tabular}
\end{table}

We have also discarded those users whose profile pictures do not have a recognizable face or have more than one recognizable face, according to Face++. Our final dataset contains $1,670,863$ users located in U.S. with identified demographic information. The phases of our data crawling and the amount of data discarded on each step are summarized in Table~\ref{table:dataset}. 

\subsection{Baseline Dataset} \label{sec:baseline}

In this section, we use the null model as our approach to estimate the statistical significance of the observed trend in given data. We compare the distribution of random samples created by the null model with the one of the original dataset and we measure the statistical significance.

Table \ref{table:expected16m} shows the distribution of gender and race in the dataset of the $\approx 1.6$ million Twitter users between July and September 2016. To construct a null model, we create $k$ random samples from the entire dataset (our crawled dataset containing 1.6 million users with demographic attributes), where each sample has exactly $304,477$ users. We choose this value for each sample size as it corresponds to the number of users we were able to gather tweets. For each sample, we count how many Whites are included. Then, the $Z_{White}$ is computed as following:

\begin{equation}
    Z_{White}=\frac{|U_{White}|-mean(|S_{White}|)}{std(|S_{White}|)}
\end{equation}

where $mean(\cdot)$ is the mean and $std(\cdot)$ is the standard deviation of the values from multiple samples. We use the same equation for the other gender and race attributes. Table \ref{table:expected} presents the demographic distribution of $304,477$ users with linguistic attributes. The numbers in the parenthesis correspond the $Z$-values.

Intuitively, when the absolute value of $Z$-value becomes bigger (either positive or negative), the trend (more number or less number, respectively) is less likely observed by chance.  In this work, we use $k$=100.

\begin{table}[t]
\centering
\caption{Demographic distribution of $1.6$ million users, our Baseline dataset.}
\label{table:expected16m}
\begin{tabular}{|c|c|c|c|}
\hline
\multirow{2}{*}{\textbf{Race (\%)}} & \multicolumn{2}{c|}{\textbf{Gender (\%)}} & \multirow{2}{*}{\textbf{Total}} \\ \cline{2-3}
                               & \textbf{Male}     & \textbf{Female}  &                                 \\ \hline
Asian                          & $7.24$   & $10.61$ & $17.85$                \\ \hline
Black                          & $7.84$   & $6.45$  & $14.29$                \\ \hline
White                          & $32.23$  & $35.63$ & $67.86$               \\ \hline
\textbf{Total}                 & $47.31$  & $52.69$ & $100.00$                 \\ \hline
\end{tabular}
\end{table}

\begin{table}[t]
\centering
\caption{Demographic distribution of $304,477$ users with linguistic attributes. The numbers in the parenthesis correspond the $Z$-values.}
\label{table:expected}
\begin{tabular}{|c|c|c|c|}
\hline
\multirow{2}{*}{\textbf{Race (\%)}} & \multicolumn{2}{c|}{\textbf{Gender (\%)}} & \multirow{2}{*}{\textbf{Total}} \\ \cline{2-3}
                               & \textbf{Male}     & \textbf{Female}  &                                 \\ \hline
Asian                          & $7.07$ ($-3.85$)   & $10.05$ ($-11.28$) & $17.12$ ($-10.90$)                \\ \hline
Black                          & $8.17$ ($8.53$)   & $6.74$ ($7.68$)  & $14.91$ ($11.69$)               \\ \hline
White                          & $32.88$ ($8.49$)  & $35.09$ ($-7.69$) & $67.97$ ($1.20$)              \\ \hline
\textbf{Total}                 & $48.12$ ($10.91$)  & $51.88$ ($-10.91$) & $100.00$                 \\ \hline
\end{tabular}
\end{table}






\subsection{Gathering Tweets} \label{subsec:followers}

We are interested in correlating linguistic features of Twitter users with demographic information. We crawled the recent $3,200$ tweets of $304,477$ users for the purpose of linguistic analysis. Table~\ref{table:expected} shows the demographic breakdown of users in our dataset across the different demographic groups. We can note a prevalence of females ($51.88\%$) in comparison to males ($48.12\%$) and a predominance of Whites ($67.97\%$) in comparison to Blacks ($14.91\%$) and Asians ($17.12\%$). This means if we pick users randomly in our dataset, we would expect demographic groups with these proportions. Table~\ref{table:tweets_stats} shows the statistical descriptions of number of tweets with $95\%$ confidence level for each demographic group.

\begin{table}[]
\centering
\caption{Basic statistical descriptions of number of tweets with confidence intervals of $95\%$ confidence level.}
\label{table:tweets_stats}
\begin{tabular}{@{}crrr@{}}
\toprule
Demographic & \multicolumn{1}{c}{Mean} & \multicolumn{1}{c}{Median} & \multicolumn{1}{c}{Max} \\ \midrule
Male         & $11,624.76 \pm 109.40$                    & $3,874$      & $1,683,948$                 \\
Female       & $12,933.40 \pm 105.89$                    & $4,885$      & $1,132,964$                 \\
Asian       & $14,020.92 \pm 183.73$                    & $5,544$      & $1,108,525$                 \\
Black       & $18,949.91 \pm 248.46$                    & $8,245$      & $973,225$                 \\
White       & $10,432.49 \pm 85.28$                     & $3,637$       & $1,683,948$                 \\ \bottomrule
\end{tabular}
\end{table}

\subsection{Extraction of Topics}

We extracted the information about topics of interests for active users using the \emph{Who Likes What}\footnote{\url{http://twitter-app.mpi-sws.org/who-likes-what}} web service \cite{Bhattacharya2014}. The produced topics are derived from the list of the friends (other users the user is following) of each user. Then, we sort the produced topics based on their frequency to conclude the $20$ most common topics from the Twitter users, including them as Binary variables. We manually cleaned several top topic labels following the same procedure as \cite{nilizadeh2016twitter}. Therefore, we merged topics like \textit{businesses} and \textit{biz}, group topics into similarity (e.g. \textit{celebrities} and \textit{famous}, \textit{actors} and \textit{actor}), and remove some topics like \textit{best}, \textit{br}, \textit{bro}, \textit{new}. Table \ref{table:topics} presents a list of the 20-top topics and the merged sub-topics in each one as well as the number of users that belong to them.

\begin{table*}[]
\centering
\caption{20-top Topics of user's interests}
\label{table:topics}
\begin{tabulary}{1\textwidth}{lLl}

\textbf{Topic} & \multicolumn{1}{c}{\textbf{Sub-Topics}}                                                                                                              & \textbf{Total} \\ \midrule
Celebrities    & celebrities, famous, stars, celebs, celebrity, star, celeb                                                                                                & $1,319,765$        \\
Artists        & musicians, singers, artist, singer, musician, rappers, bands                                                                                              & $731,370$         \\
World          & world, earth, hollywood, usa, canada, texas, international, nyc, country, city, boston, san francisco, france, america, los angeles, brasil, london, india           & $654,555$         \\
Music          & music, pop, hip hop, rap, gospel, hiphop                                                                                                                 & $463,451$         \\
Fun            & fun, funny, humor, lol, laugh                                                                                                                           & $415,113$         \\
Entertainment  & entertainment                                                                                                                                       & $371,503$         \\
TV             & tv, television                                                                                                                                       & $369,440$         \\
Info           & info, information                                                                                                                                    & $297,705$         \\
Sports         & sports, football, basketball, baseball, soccer, futbol, basket, martial arts, sport, mma, golf, cricket, boxing, motorsports, f1, racing                           & $296,652$         \\
Media          & sports news, tech news, newspapers, music news, breaking news, world news, news media, radio, internet, social media, youtube, sports media, magazines, magazine & $293,206$         \\
Life           & life, lifestyle, health, healthcare, fitness, food, style, smile, drink                                                                                     & $278,348$         \\
Actors         & actors, actresses, actress, actor                                                                                                                      & $267,626$         \\
Bloggers       & bloggers, blogs, blog                                                                                                                                 & $230,347$         \\
Technology     & technology, tech, iphone, digital, geek, software, computer, electronic, android, xbox, mac, gadgets, programming, geeks                                         & $208,739$         \\
Movie          & movie, movies, film, films                                                                                                                             & $203,577$         \\
Writers        & writers                                                                                                                                             & 189526         \\
Organizations  & organizations, nfl, nba, mlb, nhl, ufc, lfc, lgbt                                                                                                          & $178,030$         \\
Business       & business, biz, businesses                                                                                                                             & $171,759$         \\
Politics       & politics, government, political, politicians, politician                                                                                                & $110,367$         \\
Companies      & companies, apple, company, microsoft, google                                                                                                            & $79,528$          \\ \bottomrule
\end{tabulary}
\end{table*}

\subsection{Linguistic Measures}
To quantify gender and race dimensions in the language
of Twitter users, we use the 2015 version of the psycholinguistic lexicon Linguistic Inquiry and Word Count (LIWC) \cite{tausczik2010psychological}.
Since LIWC has been proposed, it has been widely used for a number of different tasks, including 
sentiment analysis~\cite{Ribeiro2016} and discourse characterization in social media platforms~\cite{correa-2015-anonymityShades}.  The features are categorized into 3 main categories, (1) affective attributes, (2) cognitive attributes, and (3) linguistic style attribute as Choudhury \textit{et al.}~\cite{DeChoudhury:2017:GCD:2998181.2998220} propose. For this work, we considered 36 features from LIWC categorized into 6 groups in order to find the main differences across each demographic group. 

The affective attributes contemplate features that show how strong is the expression of feelings like anger, anxiety, sadness, and swear. Cognitive attributes are related to the process of knowledge acquisition through perception. The lexical density and awareness group gather features related to the language itself and its structure. Temporal references are related to the tense expressed in the writing, while interpersonal focuses in present features related to the speech. The social/personal concerns group comprises features that express characteristics inherent to the individual as well his/her relation to the environment where he/she lives.

\subsection{Data Limitations}

The gender and race inference are challenge tasks, and as other existing strategies have limitations and the accuracy of Face++ inferences is an obvious concern in our effort. 
Face++ itself returns the confidence levels for the inferred gender and race attributes, and it returns an error range for inferred age. In our data, the average confidence level reported by Face++ is $95.22\pm0.015\%$ for gender and $85.97\pm0.024\%$ for race, with a confidence interval of $95\%$. Recent efforts have used Face++ for similar tasks and reported fairly well confidence in manual inspections~\cite{an2016greysanatomy,Zagheni2014,chakraborty2017@icwsm}. 
Our dataset may contain fake accounts and bots as previous studies provide evidence for a non-negligible  rate of fake accounts \cite{Freitas2015@asonam,messias13@firstmonday} in Twitter. 

Finally, we note that our approach to identify users located in U.S. may bring together some users located in the same time zone, but from different countries. We, however, believe that these users might represent a small fraction of the users, given the predominance of active U.S. users in Twitter~\cite{sysomos}.  
\section{Linguistic Differences} \label{sec:linguistics}

In order to show how demographic groups differ from each other in both gender and race domains, this section presents the difference between demographic groups across various linguistic categories. Table \ref{table:liwc_gender} shows the linguistic features extracted from LIWC into $6$ categories (affective attributes, cognitive attributes, lexical density and awareness, temporal references, social and personal concerns, and interpersonal focus).

\begin{figure}[!htb]
  \centering
    \includegraphics[width=0.49\textwidth]{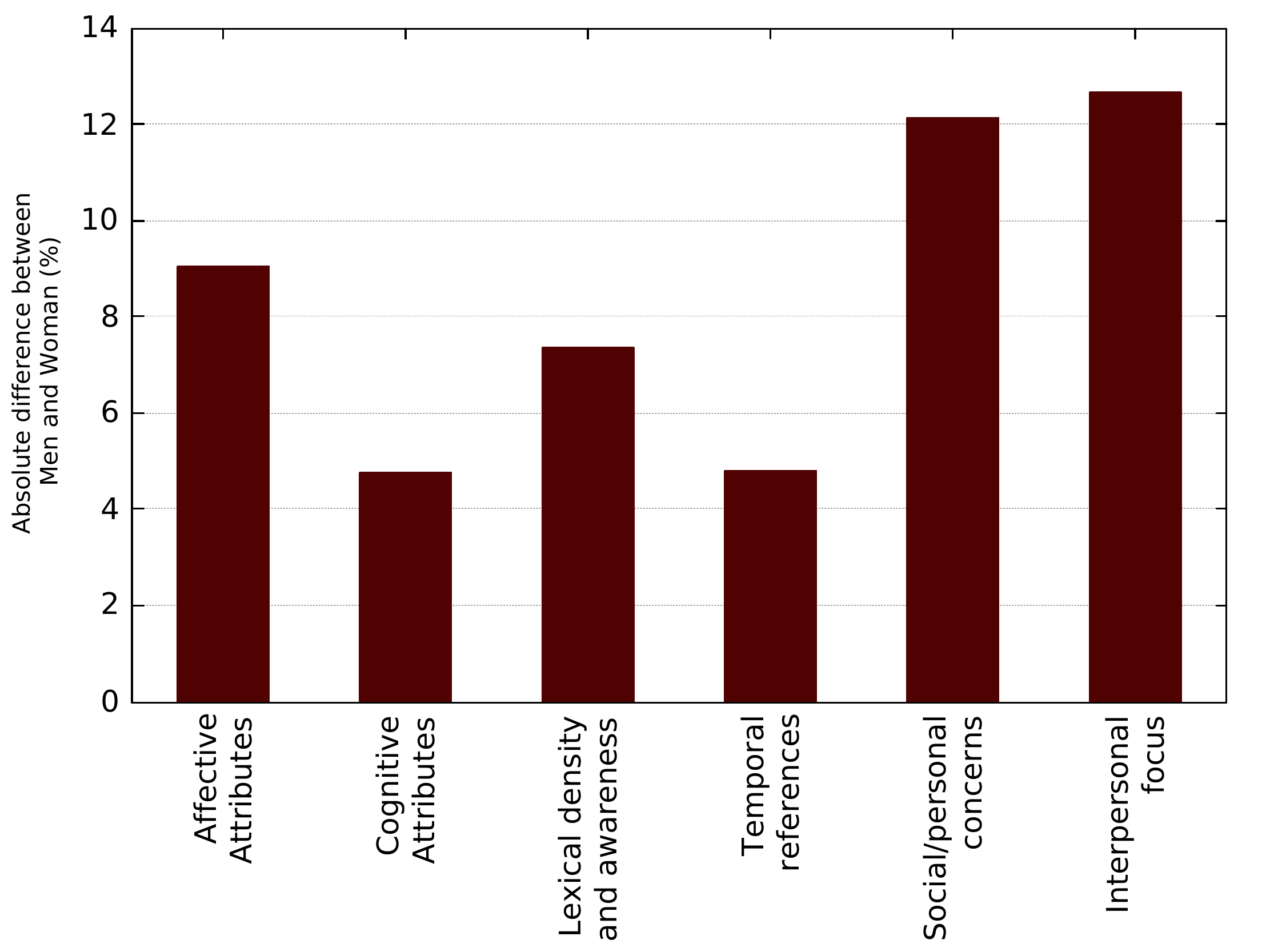}
  \caption{Mean absolute differences between male and female users per the various categories of linguistic measures}
  \label{fig:gender_abs_diff}
\end{figure}

Figure \ref{fig:gender_abs_diff} shows the mean absolute differences between male and female users across each linguistic category. The difference for a specific group of features is calculated by taking the average ratio of the difference between the values for male and female to the values of the measure among male. The mean difference in the first group (affective attributes) for instance is calculated as the average of the absolute difference of each feature that comprises this group. This shows in which linguistic categories the analyzed users differ the most. The amount of users considered in each group were the same. 

Figure \ref{fig:gender_abs_diff} also shows that interpersonal focus, which contemplates features like family, friends, health, religion, body, achievement, home, and sexual as the most prominent linguistic difference among males and females. In counterpart, from the race domain, the differences tend to be higher in affective attributes. 

\begin{figure}[!htb]
  \centering
    \includegraphics[width=0.49\textwidth]{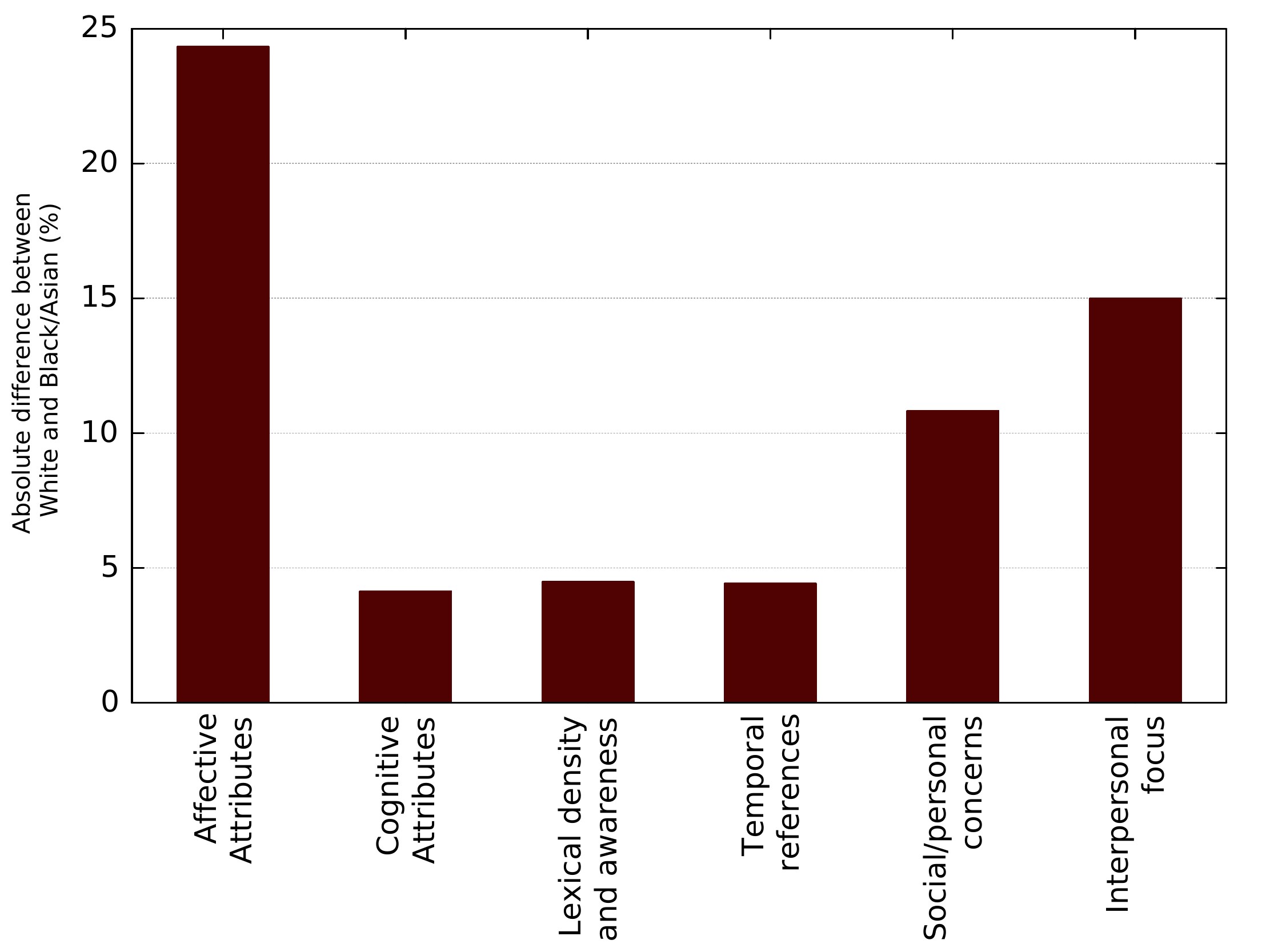}
  \caption{Mean absolute differences between White and Black/Asian users combined per the various categories of linguistic measures}
  \label{fig:white_abs_diff}
\end{figure}

In the race domain, the analysis of the linguistic difference for each race was performed in the same way as gender, but considering the other two races combined. Figure \ref{fig:white_abs_diff} shows the mean absolute differences between White and Black/Asians combined. As we can see, there is a stronger difference in affective attributes, which comprises the expression of anger, anxiety, sadness, and swear. Other linguistic aspects such as social/personal concerns and interpersonal focus showed to be relevant when comparing the writing of White users against the Black and Asian group.

Respectively, the linguistic difference among Black users was compared against White and Asian users combined. Again, affective attributes are the linguistic group with the features that most differ from one ethnicity to the others. 

\begin{figure}[!htb]
  \centering
    \includegraphics[width=0.49\textwidth]{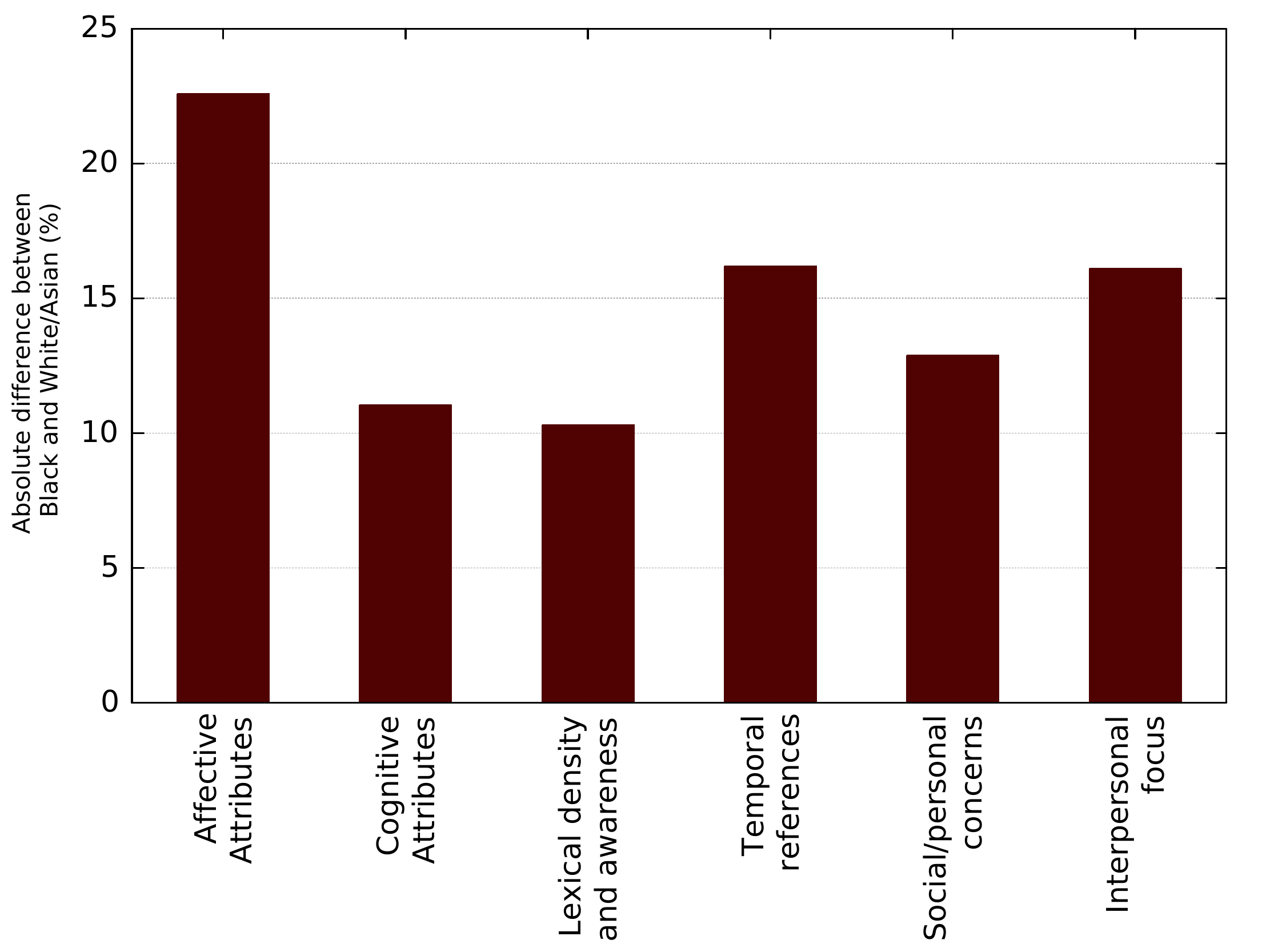}
  \caption{Mean absolute differences between Black and White/Asian users combined per the various categories of linguistic measures}
  \label{fig:black_abs_diff}
\end{figure}

When it comes to comparing the Asian linguistic to that in White and Black users, some group of features that did not present higher absolute differences when comparing Black and White groups, now tend to be higher such as lexical density and awareness and temporal references, which reveal some differences reflected by such different cultures especially in their way of writing.

\begin{figure}[!htb]
  \centering
    \includegraphics[width=0.49\textwidth]{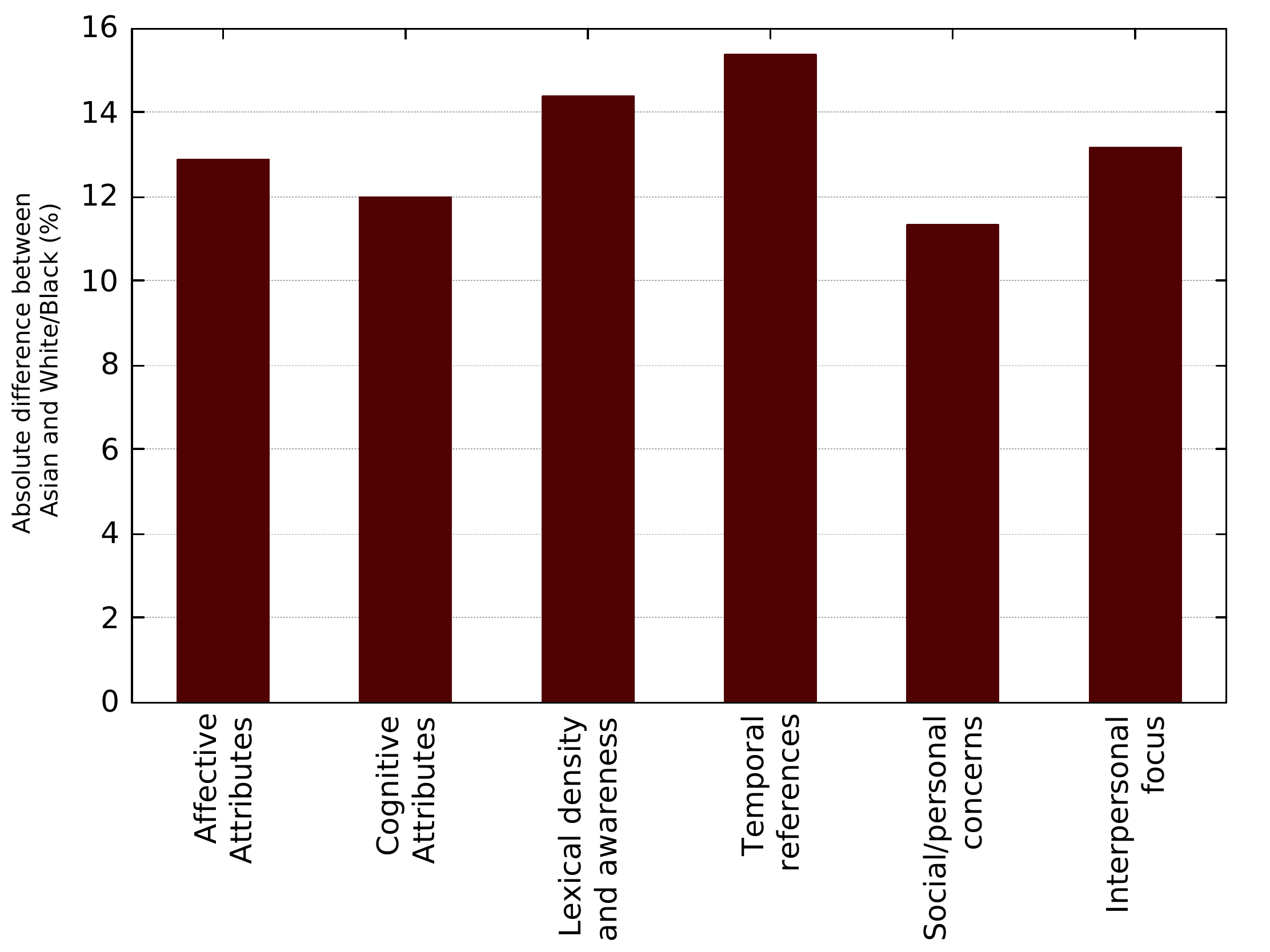}
  \caption{Mean absolute differences between Asian and White/Black users combined per the various categories of linguistic measures}
  \label{fig:asian_abs_diff}
\end{figure}

Additionally, we correlate the produced linguistic features with gender based on Wilcoxon rank sum significance tests.
$p$-values are represented in asterisks scale using * as significant ($0.1<p\leq0.5$), ** very significant($0.001<p\leq0.01$), ***($0.0001<p\leq0.001$) and ****($p<0.001$) extremely significant.
As Table \ref{table:liwc_gender} presents, females tend to use anxiety ($z=-74.534$) and sadness ($z=-74.394$) terms and phrases. On the other hand, males express with anger ($z=4.733$) in their tweets.

In terms of cognitive attributes, females are more likely to write phrases that express cognition and perception. From this group of features, two stand out: certainty ($z=-60.593$) and feel ($z=-70.766$) showing how females express more confidence and feelings in their writing.

In Lexical Density and Awareness, we can see that females make more use of verbs ($z=-45.808$), auxiliary verbs ($z=-46.441$), conjunctions ($z=-72.098$), and adverbs ($z=-66.915$), while males use more articles ($z=77.303$) and prepositions ($z=32.596$).

The temporal references attributes are more present in the females writing, as we can see from the values for present tense ($z=-62.110$) and future tense ($z=-15.118$)

From Social/Personal Concerns perspective there is a clear trend on the usage of these features by females more than by males. Among the most notorious values shown in Table \ref{table:liwc_gender}, are family ($z=-93.252$), bio ($z=-102.681$). Also, the predominance of features like friends, social, health, and body show that females express more social and personal concerns in their writing than males. The only feature in this group that is more present in males' writing is achievement ($z=65.265$)

Noticeably, females also have a higher tendency to write in the first person singular ($z=-97.329$) and in the second person ($z=-88.482$) than males, while there is a slight trend towards males using the first person plural in detriment of females ($z=4.309$).

Also, from the race perspective, the difference of values between each race shows some particularities in the way of writing for each race. In this analysis, one race is compared with the other two combined (e.g. White users are compared with Blacks and Asians).
From affective attributes, it is possible to see that Black users tend to express more anger ($z=94.610$) and swear ($z=107.344$) than White/Asian.

From cognitive attributes, almost all features were more present in Black users' texts than in the other races, with higher values for certainty ($z=62.239$), hear ($z=62.137$), and feel ($z=63.963$).

In terms of lexical density and awareness, Black users have more presence in features like verbs, auxiliary verbs, conjunctions, and adverbs, while prepositions are more present among White users. 

When talking about Social/Personal concerns, there is a higher presence of Black people in the features from this class, noticeably in family ($z=86.721$), social ($z=90.830$), religion ($z=85.163$), and body ($z=86.903$). 

The Interpersonal Focus feature set reveal that there is a predominance in the use of first person plural for White ($z=77.425$) while first person singular ($z=63.492$), second person ($z=95.495$) and third person ($z=87.717$) are more prominent in the Black group.

\begin{table}[!htb]
\centering
\caption{Differences between tweets from male and female users based on linguistic measures.
 $\mu(male)$ and $\mu(female)$ are the median values of feature for male and female, respectively. Statistical significance is count based on Wilcoxon rank sum tests.  $p$-values are represented in asterisks scale using * as significant ($0.1<p\leq0.5$), ** very significant ($0.001<p\leq0.01$), ***($0.0001<p\leq0.001$) and ****($p<0.001$) extremely significant.
}
\label{table:liwc_gender}
\begin{tabular}{lllll}
\hline
\multicolumn{1}{l|}{}                 & $\mu(male)$ & $\mu(female)$ & z \\ \hline
\multicolumn{4}{l}{\textbf{Affective attributes}}         \\ \hline
\multicolumn{1}{l|}{anger}   &     0.0055   &    0.0056   &    4.733  \\
\multicolumn{1}{l|}{anxiety} &    0.0016   &    0.0019   &    -74.534  \\
\multicolumn{1}{l|}{sadness}    &    0.0029   &    0.0034   &    -74.394 \\ 
\multicolumn{1}{l|}{swear}   &    0.0023   &    0.0026   &    -7.411 \\ \hline
\multicolumn{4}{l}{\textbf{Cognitive attributes}}         \\ \hline
\multicolumn{4}{l}{Cognition}                             \\ \hline
\multicolumn{1}{l|}{causation} &    0.0101   &    0.0104   &    -18.627 \\
\multicolumn{1}{l|}{certainty}   &    0.0101   &    0.0111   &    -60.593 \\
\multicolumn{1}{l|}{tentativeness}   &    0.0136   &    0.0141   &    -14.641  \\ \hline
\multicolumn{4}{l}{Perception}                            \\ \hline
\multicolumn{1}{l|}{see}      &    0.00957   &    0.0099   &    -24.538   \\
\multicolumn{1}{l|}{hear}         &    0.0055   &    0.0056   &    -0.033$^{*}$ \\
\multicolumn{1}{l|}{feel}    &    0.0035   &    0.0041  &    -70.766   \\
\multicolumn{1}{l|}{percepts}   &    0.0207   &    0.0218   &    -41.373     \\
\multicolumn{1}{l|}{insight}     &    0.0115   &    0.0125   &    -46.806    \\
\multicolumn{1}{l|}{relative}    &    0.1014   &    0.0999   &    18.026  \\ \hline
\multicolumn{4}{l}{\textbf{Lexical Density and Awareness}} \\ \hline
\multicolumn{1}{l|}{verbs}  &    0.1103   &    0.1170   &    -45.808    \\
\multicolumn{1}{l|}{auxiliary verbs}  &    0.0539   &    0.0583   &    -46.441    \\
\multicolumn{1}{l|}{articles}   &    0.0370   &    0.0340   &    77.303       \\
\multicolumn{1}{l|}{prepositions}    &    0.0843   &    0.0817   &    32.596       \\
\multicolumn{1}{l|}{conjunctions}  &    0.0279   &    0.0314   &    -72.098        \\
\multicolumn{1}{l|}{adverbs}    &    0.0317   &    0.0355   &    -66.915       \\ \hline
\multicolumn{4}{l}{\textbf{Temporal references}}          \\ \hline
\multicolumn{1}{l|}{present tense}  &    0.0802   &    0.0871   &    -62.110        \\
\multicolumn{1}{l|}{future tense}    &    0.0103   &    0.0106   &    -15.118     \\ \hline
\multicolumn{4}{l}{\textbf{Social/Personal Concerns}}     \\ \hline
\multicolumn{1}{l|}{family}        &    0.0026   &    0.0034   &    -93.252    \\
\multicolumn{1}{l|}{friends}       &    0.0028   &    0.0033   &    -66.168    \\
\multicolumn{1}{l|}{social}      &    0.0938   &    0.1021   &    -77.896   \\
\multicolumn{1}{l|}{health}     &    0.0037   &    0.0044   &    -76.446     \\
\multicolumn{1}{l|}{religion}   &    0.0024   &    0.0025   &    -26.485       \\
\multicolumn{1}{l|}{bio}       &    0.0157   &    0.0203   &    -102.681    \\
\multicolumn{1}{l|}{body}       &    0.0045   &    0.0056   &    -58.386    \\
\multicolumn{1}{l|}{achievement}   &    0.0116   &    0.0105   &    65.265     \\
\multicolumn{1}{l|}{home}    &    0.0022   &    0.0026   &    -74.049      \\
\multicolumn{1}{l|}{sexual}   &    0.0011   &    0.0012   &    -18.691     \\
\multicolumn{1}{l|}{death}  &    0.0014   &    0.0013   &    29.463     \\ \hline
\multicolumn{4}{l}{\textbf{Interpersonal focus}}          \\ \hline
\multicolumn{1}{l|}{1st p. singular}  &    0.0245   &    0.0340   &    -97.329   \\
\multicolumn{1}{l|}{1st p. plural}   &    0.0046   &    0.0045   &    4.309    \\
\multicolumn{1}{l|}{2nd p.}   &    0.0160   &    0.0198   &    -88.482    \\
\multicolumn{1}{l|}{3rd p.}   &    0.0030   &    0.0031   &    -3.371$^{***}$  \\ \hline
\end{tabular}
\end{table}

\begin{table*}[!htb]
\centering
\caption{Differences between tweets from White, Black, and Asian users based on linguistic measures.
$\mu(White)$, $\mu(Black)$ and $\mu(Black)$ is the median value of features for each demographic group respectively.
Statistical significance is count based on Wilcoxon rank sum tests. The $p$-values present extremely significant for all linguistic features. We test the correlation of each unique demographic group with the others.}
\label{table:liwc_race}
\begin{tabular}{llllllll}
\hline
\multicolumn{1}{l|}{}                 & $\mu(White)$ & $\mu(Black)$ & $\mu(Asian)$ & $z_{W/B-A}$ & $z_{B/W-A}$ & $z_{A/W-B}$\\ \hline
\multicolumn{5}{l}{\textbf{Affective attributes}}         &    &    &    \\ \hline
\multicolumn{1}{l|}{anger}   &   0.0051   &   0.0081   &   0.0056   &   -67.261   &   94.610   &   -5.236  \\
\multicolumn{1}{l|}{anxiety}   &   0.0017   &   0.0019   &   0.0016   &   -0.696   &   33.789   &   -30.517  \\
\multicolumn{1}{l|}{sadness}   &   0.0031   &   0.0034   &   0.0032   &   -20.814   &   28.205   &   -0.625  \\
\multicolumn{1}{l|}{swear}   &   0.0021   &   0.0064   &   0.0027   &   -90.375   &   107.344   &   11.329  \\ \hline
\multicolumn{5}{l}{\textbf{Cognitive attributes}}         &    &    &    \\ \hline
\multicolumn{5}{l}{Cognition}                             &    &    &    \\ \hline
\multicolumn{1}{l|}{causation}   &   0.0104   &   0.0105   &   0.0096   &   29.931   &   19.465   &   -54.832  \\
\multicolumn{1}{l|}{certainty}   &   0.0105   &   0.0116   &   0.0101   &   -19.404   &   62.239   &   -33.955  \\
\multicolumn{1}{l|}{tentativeness}   &   0.0138   &   0.0152   &   0.0130   &   -8.958   &   55.174   &   -40.226  \\ \hline
\multicolumn{5}{l}{Perception}                            &    &        \\ \hline
\multicolumn{1}{l|}{see}   &   0.0098   &   0.0098   &   0.0095   &   18.756   &   6.970   &   -29.506  \\
\multicolumn{1}{l|}{hear}   &   0.0055   &   0.0062   &   0.0054   &   -26.349   &   62.137   &   -25.331  \\
\multicolumn{1}{l|}{feel}   &   0.0037   &   0.0044   &   0.0039   &   -44.180   &   63.963   &   -5.128  \\
\multicolumn{1}{l|}{percepts}   &   0.0212   &   0.0223   &   0.0210   &   -14.067   &   43.711   &   -23.308  \\
\multicolumn{1}{l|}{insight}   &   0.0122   &   0.0128   &   0.0112   &   11.133   &   40.420   &   -51.201  \\
\multicolumn{1}{l|}{relative}   &   0.1020   &   0.1012   &   0.0936   &   50.614   &   15.841   &   -76.870   \\ \hline
\multicolumn{5}{l}{\textbf{Lexical Density and Awareness}} &    &    &    \\ \hline
\multicolumn{1}{l|}{verbs}   &   0.1125   &   0.1222   &   0.1082   &   -16.435   &   64.214   &   -39.436  \\
\multicolumn{1}{l|}{auxiliary verbs}   &   0.0554   &   0.0612   &   0.0529   &   -12.202   &   58.285   &   -39.130  \\
\multicolumn{1}{l|}{articles}   &   0.0366   &   0.0339   &   0.0314   &   96.532   &   -26.056   &   -94.363  \\
\multicolumn{1}{l|}{prepositions}   &   0.0851   &   0.0817   &   0.0743   &   77.024   &   1.032   &   -95.556  \\
\multicolumn{1}{l|}{conjunctions}   &   0.0291   &   0.0319   &   0.0286   &   -11.852   &   43.571   &   -25.898  \\
\multicolumn{1}{l|}{adverbs}   &   0.0329   &   0.0363   &   0.0325   &   -17.239   &   48.159   &   -23.542  \\ \hline
\multicolumn{5}{l}{\textbf{Temporal references}}          &    &    &    \\ \hline
\multicolumn{1}{l|}{present tense}   &   0.0825   &   0.0912   &   0.0798   &   -21.972   &   69.126   &   -37.196  \\
\multicolumn{1}{l|}{future tense}   &   0.0103   &   0.0119   &   0.0099   &   -28.333   &   79.181   &   -38.719     \\ \hline
\multicolumn{5}{l}{\textbf{Social/Personal Concerns}}     &    &    &    \\ \hline
\multicolumn{1}{l|}{family}   &   0.0029   &   0.0040   &   0.0032   &   -74.318   &   86.721   &   10.755  \\
\multicolumn{1}{l|}{friend}   &   0.0031   &   0.0033   &   0.0033   &   -26.248   &   25.332   &   8.717  \\
\multicolumn{1}{l|}{social}   &   0.0956   &   0.1101   &   0.0971   &   -60.389   &   90.830   &   -10.166  \\
\multicolumn{1}{l|}{health}   &   0.0040   &   0.0044   &   0.0039   &   -9.579   &   45.973   &   -30.920  \\
\multicolumn{1}{l|}{religion}   &   0.0024   &   0.0031   &   0.0024   &   -53.672   &   85.163   &   -13.154  \\
\multicolumn{1}{l|}{bio}   &   0.0176   &   0.0204   &   0.0179   &   -32.215   &   53.914   &   -10.492  \\
\multicolumn{1}{l|}{body}   &   0.0048   &   0.0067   &   0.0052   &   -62.906   &   86.903   &   -3.428  \\
\multicolumn{1}{l|}{achievement}   &   0.0114   &   0.0109   &   0.0097   &   69.227   &   -1.632   &   -83.506  \\
\multicolumn{1}{l|}{home}   &   0.0025   &   0.0024   &   0.0022   &   50.362   &   -4.554   &   -57.624  \\
\multicolumn{1}{l|}{sexual}   &   0.0011   &   0.0019   &   0.0012   &   -51.768   &   71.799   &   -3.084  \\
\multicolumn{1}{l|}{death}  &   0.0014   &   0.0015   &   0.0013   &   4.356   &   31.454   &   -34.554  \\\hline
\multicolumn{5}{l}{\textbf{Interpersonal focus}}          &    &    &    \\ \hline
\multicolumn{1}{l|}{1st p. singular}  &   0.0268   &   0.0355   &   0.0296   &   -51.874   &   63.492   &   4.760  \\
\multicolumn{1}{l|}{1st p. plural}   &   0.0048   &   0.0042   &   0.0039   &   77.425   &   -28.107   &   -68.994  \\
\multicolumn{1}{l|}{2nd p.}      &   0.0169   &   0.0227   &   0.0177   &   -63.930   &   95.495   &   -10.148  \\
\multicolumn{1}{l|}{3rd p.}   &   0.0030   &   0.0039   &   0.0028   &   -36.070   &   87.717   &   -37.143  \\ \hline
\end{tabular}
\end{table*}

Table \ref{table:ranking-top-20-gender} \& Table \ref{table:ranking-top-20-race} present the ranking difference for the 20 most common phrases for gender and races respectively. To find these differences, we randomly selected $1,000$ users from each group (male, female, Asian, Black and White). Their tweets were used to create ngrams for each group. With this subset of our dataset, we extracted the top $100$ phrases for each demographic group and the top $20$ are shown in these Tables.

\begin{table}[!htb]
\centering
\caption{Ranking Differences of Gender Top Phrases. We use $ne$ for no existing phrases in a group.}
\label{table:ranking-top-20-gender}
\begin{tabular}{rccc}
\hline
\textbf{}                            & \textbf{Rank(Female)} & \textbf{Rank(Male)} & \textbf{DifF(F-M)} \\ \hline
\multicolumn{1}{r|}{i do n't}        & 1                    & 1                  & 0                  \\
\multicolumn{1}{r|}{i ca n't}        & 2                    & 2                  & 0                  \\
\multicolumn{1}{r|}{you do n't}      & 3                    & 3                  & 0                  \\
\multicolumn{1}{r|}{i 'm not}        & 4                    & 4                  & 0                  \\
\multicolumn{1}{r|}{ca n't wait}     & 5                    & 8                  & 3                  \\
\multicolumn{1}{r|}{i 'm so}         & 6                    & 19                 & 13                 \\
\multicolumn{1}{r|}{i love you}      & 7                    & 15                 & 8                  \\
\multicolumn{1}{r|}{do n't know}     & 8                    & 11                 & 3                  \\
\multicolumn{1}{r|}{i want to}       & 9                    & 24                 & 15                 \\
\multicolumn{1}{r|}{more for virgo}  & 10                   & 55                 & 45                 \\
\multicolumn{1}{r|}{more for cancer} & 11                   & 29                 & 18                 \\
\multicolumn{1}{r|}{i wan na}        & 12                   & 28                 & 16                 \\
\multicolumn{1}{r|}{! i 'm}          & 13                   & 25                 & 12                 \\
\multicolumn{1}{r|}{you ca n't}      & 14                   & 16                 & 2                  \\
\multicolumn{1}{r|}{more for libra}  & 15                   & 39                 & 24                 \\
\multicolumn{1}{r|}{it 's a}         & 16                   & 10                 & 6                  \\
\multicolumn{1}{r|}{and i 'm}        & 17                   & 33                 & 16                 \\
\multicolumn{1}{r|}{more for pisces} & 18                   & ne                 & -                  \\
\multicolumn{1}{r|}{i need to}       & 19                   & 34                 & 15                 \\
\multicolumn{1}{r|}{do n't have}     & 20                   & 27                 & 7                  \\ \hline
\end{tabular}
\end{table}

As we can see in Table \ref{table:ranking-top-20-gender} phrases expressing negation are in the top positions for both males and females. It is also clear to see that females are more into signs than males since phrases with this kind of content present higher differences in the gender ranking. 

Due to the informal nature of Twitter, the top phrases also reveal that it is common the usage of slangs like "do n't", "ca n't" and "wan na" for both genders. 

\begin{table*}[!htb]
\centering
\caption{Ranking Differences of Race Top Phrases. We use $ne$ for no existing phrases in a group.}
\label{table:ranking-top-20-race}
\begin{tabular}{r|cccccc}
\textbf{}     & \textbf{Rank(White)} & \textbf{Rank(Black)} & \textbf{Rank(Asian)} & \textbf{Diff(W-B)} & \textbf{Diff(W-A)} & \textbf{Diff(B-A)} \\ \hline
i do n't      & 1                    & 1                    & 1                    & 0         & 0         & 0         \\
i ca n't      & 2                    & 2                    & 2                    & 0         & 0         & 0         \\
ca n't wait   & 3                    & 18                   & 7                    & 15        & 4         & 11        \\
you do n't    & 4                    & 4                    & 3                    & 0         & 1         & 1         \\
i 'm not      & 5                    & 8                    & 6                    & 3         & 1         & 2         \\
i love you    & 6                    & 33                   & 4                    & 27        & 2         & 29        \\
i 'm so       & 7                    & 16                   & 6                    & 9         & 1         & 10        \\
do n't know   & 8                    & 19                   & 11                   & 11        & 3         & 8         \\
it 's a       & 9                    & 26                   & 16                   & 17        & 7         & 10        \\
one of the    & 10                   & 48                   & 20                   & 38        & 10        & 28        \\
i want to     & 11                   & 47                   & 10                   & 36        & 1         & 37        \\
! i 'm        & 12                   & 46                   & 29                   & 34        & 17        & 17        \\
if you 're    & 13                   & 28                   & 19                   & 15        & 6         & 9         \\
thank you for & 14                   & 126                  & 28                   & 112       & 14        & 98        \\
it 's not     & 15                   & 34                   & 32                   & 19        & 17        & 2         \\
and i 'm      & 16                   & 58                   & 21                   & 42        & 5         & 37        \\
you ca n't    & 17                   & 17                   & 17                   & 0         & 0         & 0         \\
i 'm at       & 18                   & 53                   & 26                   & 35        & 8         & 27        \\
n't wait to   & 19                   & 100                  & 51                   & 81        & 32        & 49        \\
i liked a     & 20                   & 7                    & ne                   & 13        & -         & -         \\ \hline
\end{tabular}
\end{table*}

When analyzing the ranking of race top phrases in Table \ref{table:ranking-top-20-race}, the trend of using negation phrases also repeat here. Phrases containing expressions like "i don't", "i can't" and "i'm not" appear in the top positions for all the racial groups. Another interesting result is the position of the expression "i love you" in the writing of different races. White and Asian users seem to be more likely to tweet contents with this expression than Black users. Also, the expression "i want to" appears more often in the writing of White and Asian users than in the Blacks. Table~\ref{table:ranking-top-20-gender} and Table~\ref{table:ranking-top-20-race} show differences regarding the way of writing of each demographic group and reveal interesting characteristics about the difference from one to another.

\section{Differences in Topic Interests} \label{sec:topics}

\begin{figure}[!htb]
  \centering
    \includegraphics[width=0.49\textwidth]{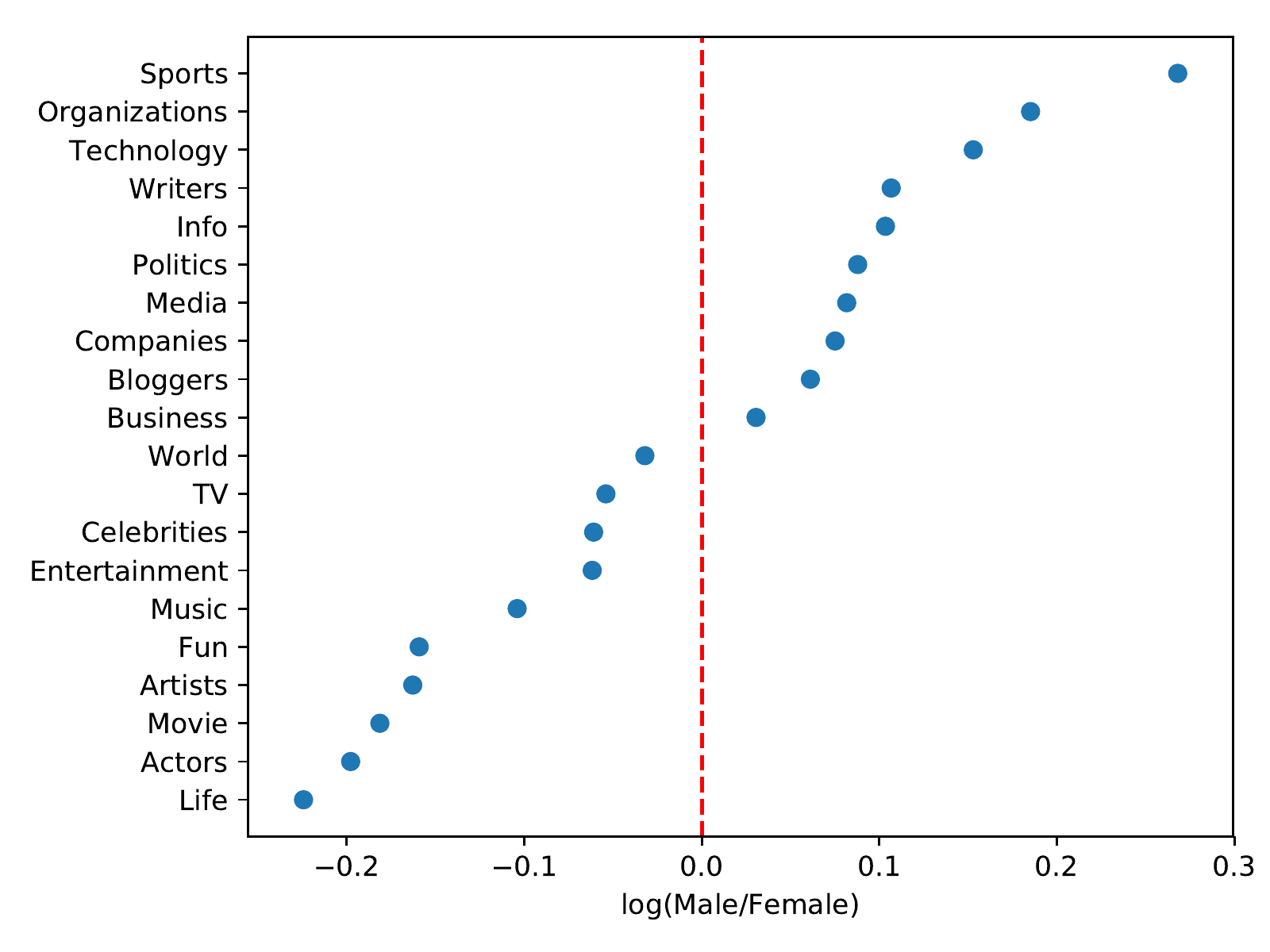}
  \caption{Gender interests: Blue dots represent the gender interests for the 20-top popular topics.}
  \label{fig:topics_gender}
\end{figure}

Males and females may have differences in preferences and interests in digest information. In order to understand which topic is preferable to females than males, we analyze the differences in the topic interest of users in our dataset. The Figure \ref{fig:topics_gender} shows the gender distribution for the 20-top topics that we extracted, with log-ratio of perceived male to female. It shows the topic interest for users based on gender in our dataset. On the right side, we see topics related to males' interests while on the left side we see the topics that females are more interested than males. The 3-top topics for males are sports, organizations, and technology. In other words, males tend to interest more in these topics than females. However, females interest more for life, actors, and movie than males. More specifically, the gender difference between topics varies among males and females.

\begin{figure*}[tb]
\center
\subfigure[White vs Asian]{
\includegraphics[width=.32\textwidth]{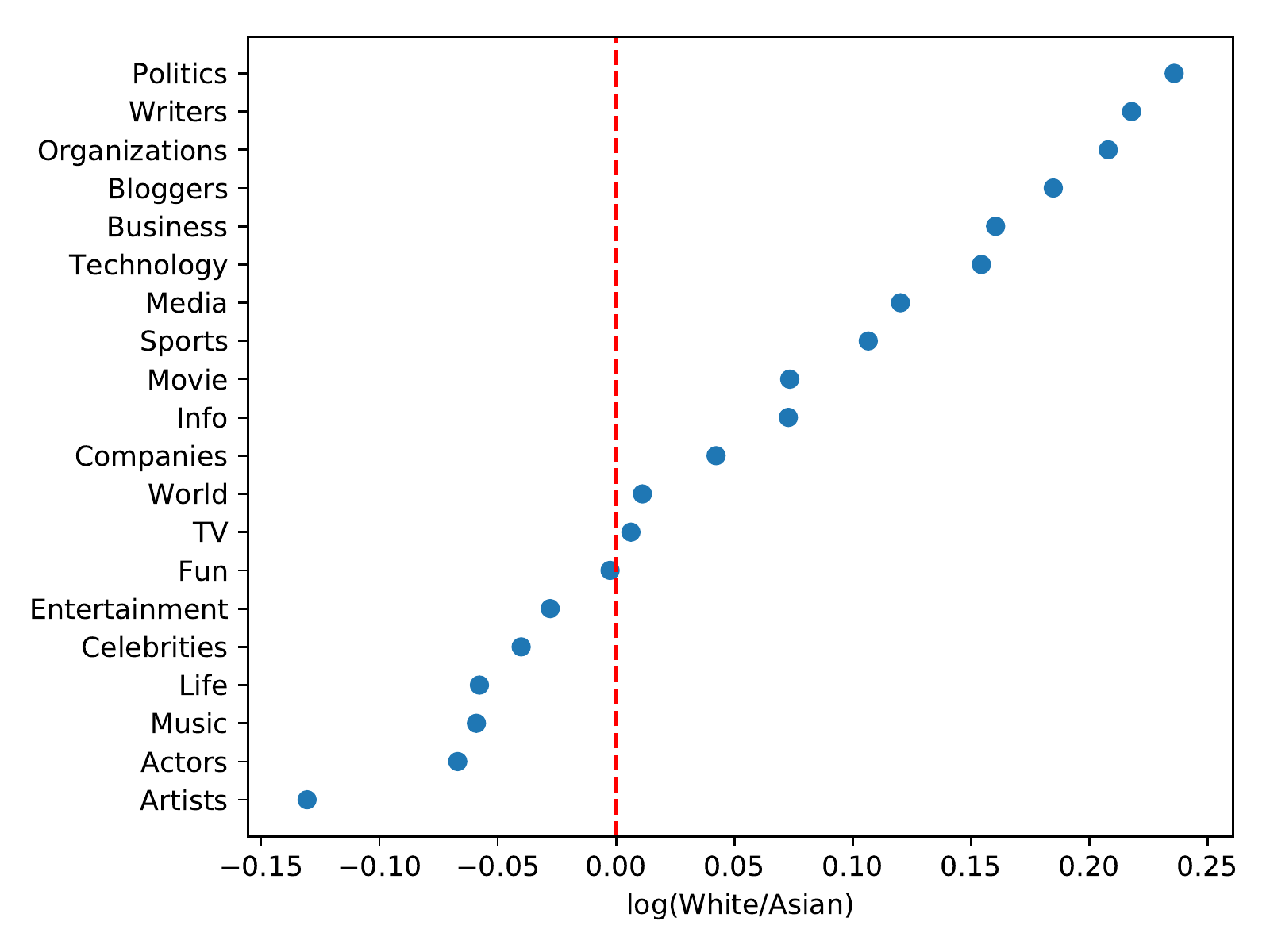}
}
\hfil
\subfigure[White vs Black]{
\includegraphics[width=.32\textwidth]{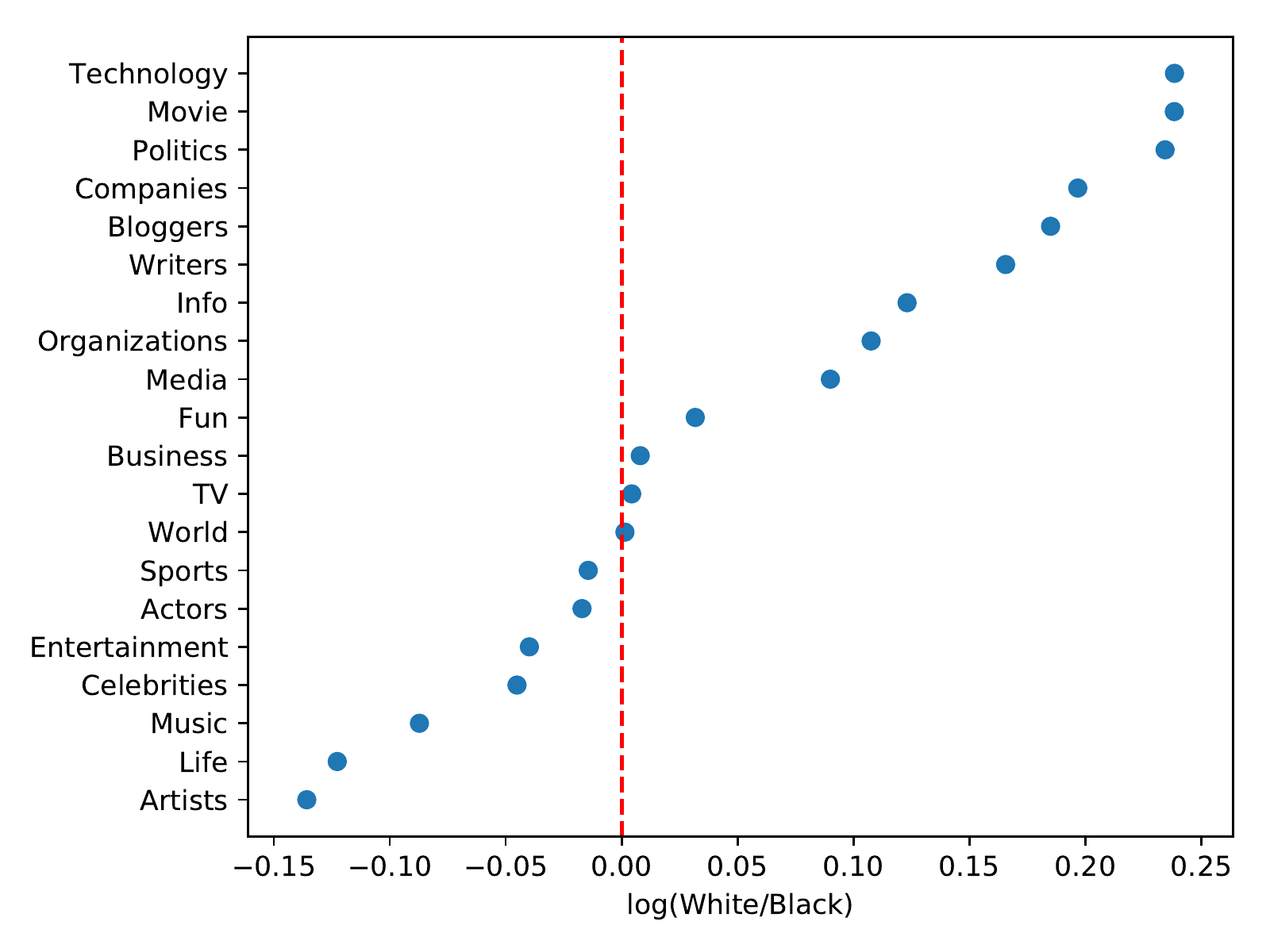}
}
\hfil
\subfigure[Asian vs Black]{
\includegraphics[width=.32\textwidth]{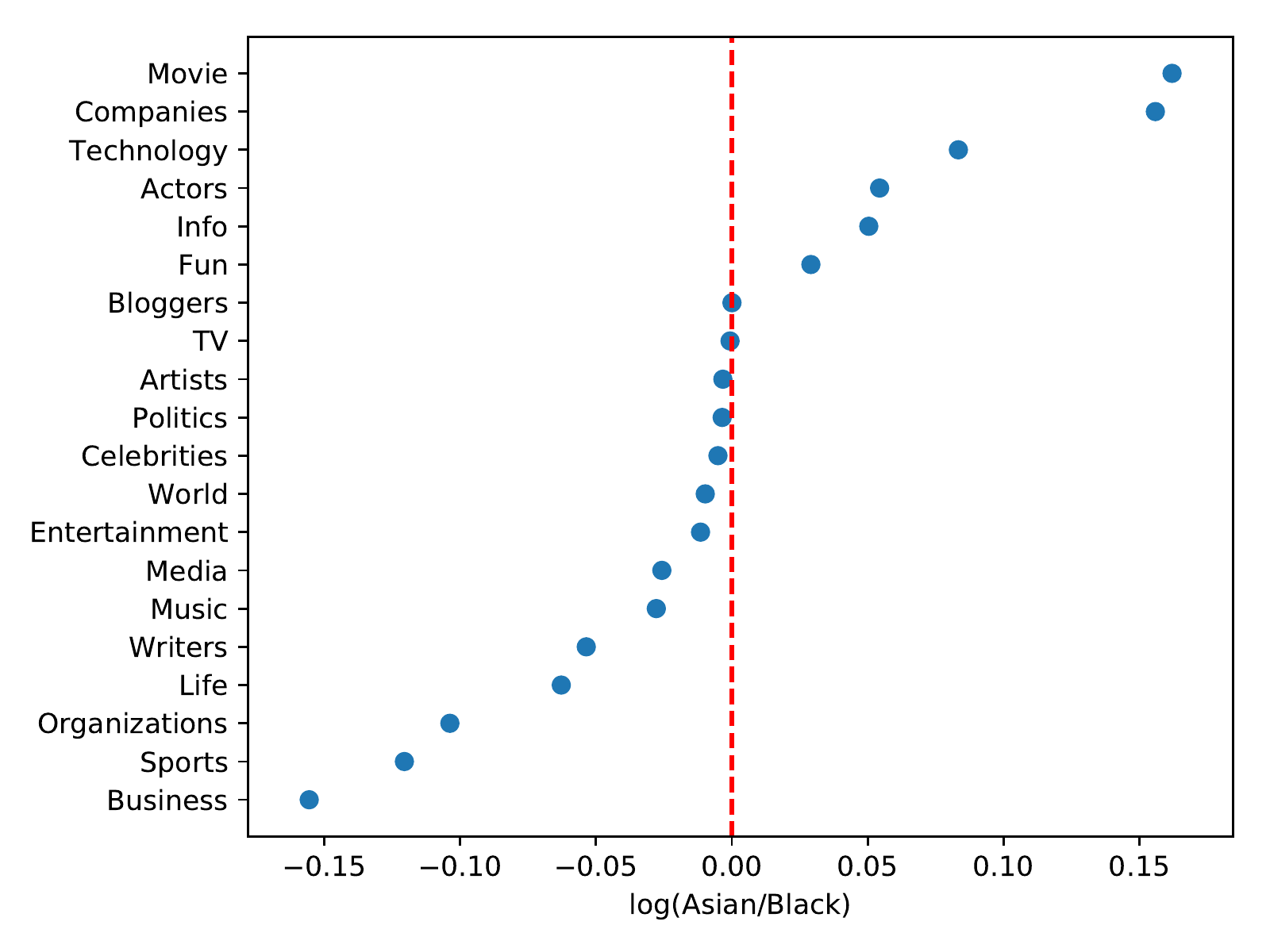}
}
\caption{Race interests: Blue dots represent the race interests of (a) White against Asians, (b) White against Blacks, and (c) Asian against Blacks for the 20-top popular topics. The dataset is normalized by the number of Blacks as shown in Table \ref{table:expected}.}
\label{fig:race_topics_against}
\end{figure*}

In a similar way, we present the race distribution for the 20-top topics of Asian, Black, and White users in Figure \ref{fig:race_topics_against}. In order to show results regarding race, for this specific analysis, we have normalized the dataset by the number of Black users once they are the minority amount of users in our dataset, as shown in Table \ref{table:expected}. Therefore, we have randomly selected $45,398$ users for each race to study their topic interests. Users from different races may also vary in interests and preferences. Figure \ref{fig:race_topics_against}-a shows that White users have more interest in politics, writers, and organizations than Asians. However, Asians prefer more artists, actors, and music topics than Whites. Figure \ref{fig:race_topics_against}-b compares the differences in topic interests for White and Blacks. We see that White users are interested in technology, movie, and politics more than Blacks. Nonetheless, Blacks prefer more artists, life, and music topics. Finally, when we look at Figure \ref{fig:race_topics_against}-c, Asians interest more for movie, companies, and technology topics than Blacks. On other hand, Blacks prefer more business, sports, and organizations than Asians.

\section{Conclusion} \label{sec:conclusion}
The results presented in this paper allow us to conclude that there are clear differences in the way of writing across different demographic groups in both gender and race domains. 
Our main contribution relies on characterizing the differences in the way of writing for each group pointing the most important linguistic aspects for a specific gender and race. Through the analysis of mean absolute differences amongst linguistic features between each demographic group, we were able to identify those which affective attributes were more present in their writing. In the same way, features based on cognitive attributes, temporal references, social and personal concerns, and interpersonal focus showed to have different weights throughout different demographic domains. 

Another interesting conclusion is based on the most common phrases encountered on each group and their position ranking when compared to different demographic groups. The analysis of these most common phrases led us to conclude that phrases expressing negation figure as one of the most frequent for all domains. Also, the usage of slangs, which is common in an environment like Twitter, appears in these frequent phrases too. When we compare the difference between the groups, we find interesting trends, like the higher interest in signs by females than by males.

By analyzing topic interests, we found that each demographic group tends to have its own preferences over the information they share. For instance, we found that males are more into sports, organizations, and technology while females have more interest in topics related to life, actors, and movie. In the same way, users from different races are also likely to have different interests and preferences. White users are more interested in politics, writers, and organizations when compared to Asians, and technology, movie, and politics when compared to Black users. On the other hand, Black users are more into artists, life, and music topics. When we look into Asians, they are more interested in artists, actors, and music than Whites and tend to have higher interest for movie, companies, and technology when compared to Blacks.

There are some future directions we would like to pursue next. 
First, we plan to study the correlation of linguistic differences with other demographic factors e.g. age. We plan to use our extracted linguistic characteristics as a feature vector for prediction of gender and race. Also, our will is to extend this work correlating demographic aspects with the social behavior, e.g. number of followers, listed, etc. In addition, we plan to examine the speed of tweets that are propagated through a specific demographic group.


\section*{Acknowledgments} \label{sec:acknowledgments}


\small{This work was partially supported by the project FAPEMIG-PRONEX-MASWeb, Models, Algorithms and Systems for the Web, process number APQ-01400-1 and grants from CNPq, CAPES, Fapemig, and Humboldt Foundation.}

\small
\bibliographystyle{SIGCHI-Reference-Format}
\bibliography{references}

\end{document}